\documentstyle[11pt]{article}
\newcount\Mac  \Mac=0 
\newcommand{\fig}[1]{~\ref{fig:#1}}
\newcommand{\GeV}{\,{\rm GeV}}

\newcommand{\TeV}{\,{\rm TeV}}
\newcommand{\dq}{{\rm d}}
\newcommand{\bAk}[3]{\langle #1|#2|#3\rangle}
\renewcommand{\Im}{\hbox{\rm Im}\,}
\renewcommand{\Re}{\hbox{Re}\,}
\newcommand{\MeV}{\,{\rm MeV}}

\def\Red{}
\def\Black{}
\def\Blue{}
\def\Green{}
\newcommand{\lascia}[1]{}
\def\puttag(#1,#2)#3{\put(#1,#2){\makebox(0,0){\rm\Blue #3\Black}}}

\def\circa#1{\,\raise.3ex\hbox{$#1$\kern-.75em\lower1ex\hbox{$\sim$}}\,}

\newcommand{\diag}{\hbox{diag}\,}
\newcommand{\ecm}{e\,{\rm cm}}

\def\putps(#1,#2)(#3,#4)#5#6{\ifnum\Mac=1 \put(#1,#2){\special{picture #5}}
\else  \put(#3,#4){\includegraphics{#6}} \fi}

\def\One{\hbox{1\kern-.24em I}}

\newcommand{\eq}[1]{~{\rm (\ref{eq:#1})}}
\newcommand{\sys}[1]{~{\rm (\ref{sys:#1})}}

\newcommand{\NP}{Nucl. Phys.}
\newcommand{\PRL}{Phys. Rev. Lett.}
\newcommand{\PL}{Phys. Lett.}
\newcommand{\PR}{Phys. Rev.}
\newcommand{\BR}{{\rm B.R.}}
\newcommand{\BRmueg}{{\rm B.R.}(\mu\to e\gamma)}
\makeatletter
\def\art{\@ifnextchar[{\eart}{\oart}}
\def\eart[#1]#2#3#4#5#6{{\rm #2}, {\em #3 \bf #4} {\rm (#6) #5} ({\em #1})}
\def\hepart[#1]#2{{\rm #2, \em#1}}
\newcommand{\oart}[5]{{\rm #1}, {\em #2 \bf #3} {\rm (#5) #4}}
\newcounter{alphaequation}[equation]
\def\thealphaequation{\theequation\hbox to
0.6em{\hfil\alph{alphaequation}\hfil}}
\def\eqnsystem#1{
\def\@eqnnum{{\rm (\thealphaequation)}}
\def\@@eqncr{\let\@tempa\relax \ifcase\@eqcnt \def\@tempa{& & &} \or
  \def\@tempa{& &}\or \def\@tempa{&}\fi\@tempa
  \if@eqnsw\@eqnnum\refstepcounter{alphaequation}\fi
\global\@eqnswtrue\global\@eqcnt=0\cr}
\refstepcounter{equation} \let\@currentlabel\theequation \def\@tempb{#1}
\ifx\@tempb\empty\else\label{#1}\fi
\refstepcounter{alphaequation}
\let\@currentlabel\thealphaequation
\global\@eqnswtrue\global\@eqcnt=0 \tabskip\@centering\let\\=\@eqncr
$$\halign to \displaywidth\bgroup \@eqnsel\hskip\@centering
$\displaystyle\tabskip\z@{##}$&\global\@eqcnt\@ne
\hskip2\arraycolsep\hfil${##}$\hfil& \global\@eqcnt\tw@\hskip2\arraycolsep
$\displaystyle\tabskip\z@{##}$\hfil
\tabskip\@centering&\llap{##}\tabskip\z@\cr}

\def\endeqnsystem{\@@eqncr\egroup$$\global\@ignoretrue} \makeatother

\begin{document}
\begin{quote}
{\em 3/8/1999}\hfill {\bf IFUP-TH/99-45}\\
hep-ph/9908255 \hfill{\bf SNS-PH/99-13}
\end{quote}
\vspace{1cm}
\begin{center}
{\LARGE\bf\Red $\epsilon'$ from supersymmetry with non universal $A$ terms?}\\[1cm]\Black
{\bf R. Barbieri, R. Contino}\\[3mm]
\normalsize\em  Scuola Normale Superiore,
Piazza dei Cavalieri 7, I-56126 Pisa, Italia\\
and INFN, Sezione di Pisa, I-56127 Pisa, Italia\\[7mm]
\Black
{\bf A. Strumia}\\[0.3cm]
\normalsize\em 
Dipartimento di Fisica, Universit\`a di Pisa and\\
INFN, Sezione di Pisa, I-56127 Pisa, Italia\\[8mm]
\Blue\large\bf Abstract
\begin{quote}\large\indent
In supersymmetric theories with a motivated flavour structure
we investigate the possibility that an $\epsilon'$ parameter as large as the
measured value be generated without conflicting with
the bounds from electron and neutron electric dipoles and $\mu\to e \gamma$.
\end{quote}
\end{center}\Black

\thispagestyle{empty}

\section{Introduction}
\begin{figure}[t]
\begin{center}
$$
\begin{picture}(17,4)
\putps(-0.5,0)(-0.5,0){fstoria}{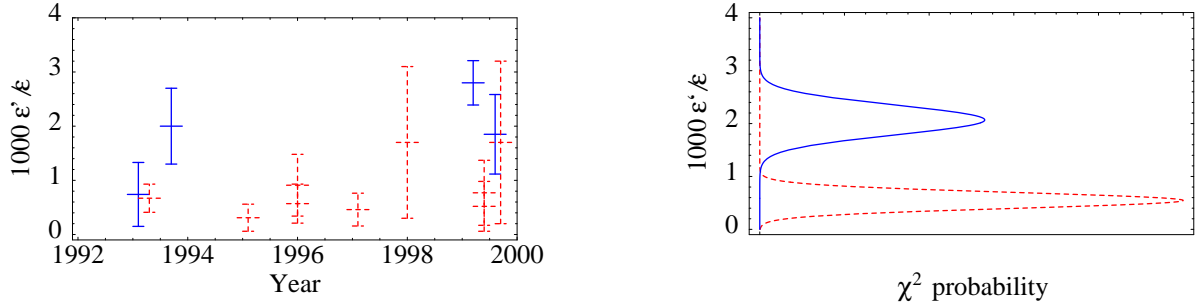}
\end{picture}$$
\caption[SP]{\em Comparison between experimental determinations (continous lines)
and theoretical predictions (dotted lines) for $\epsilon'/\epsilon$.
\label{fig:exp/th}}
\end{center}\end{figure}

Experiments~\cite{EpsPexp} have shown beyond any reasonable doubt that direct
CP violation exists in $K$ physics, as conventionally described by
the parameter $\epsilon'$, expected to be non zero in the Standard Model (SM).
A `simple minded' comparison between published theoretical predictions
for $\epsilon'$ and experimental data
shows, however,
that it is unlikely that SM physics can account for the observed value of $\epsilon'$.
In fig.\fig{exp/th}a we show the chronological history of the accurate
theoretical predictions~\cite{EpsPth} and
of the experimental data~\cite{EpsPexp}.
In fig.\fig{exp/th}b we compare the $\chi^2$-probability of the
theoretical and experimental values obtained by combining in an uncorrelated way
the data shown in
fig.\fig{exp/th}a: the probability that a fluctuation
can account for the discrepancy appears very small.
If we believe in the data that we have employed,
we are almost sure that new physics has been discovered.

Even forgetting that, for various years after 1993, the experimental value of $\epsilon'$ seemed
smaller than the `true' value found by the CERN experiment and confirmed
by 1999 results --- a fact that might have had some
psychological influence on theorists --- there is, however, at least one important reason
for being more careful.
No one of the theoretical predictions~\cite{EpsPth}, except the one with the largest error
(and the highest central value)~\cite{trieste}
is made in a framework that includes, at the same time, an explanation
of the `$\Delta I=1/2$' rule. The obvious connection between the two calculations,
although the latter may be more difficult than the former (at least for some techniques)
suggests a more conservative attitude with respect to the theoretical errors\footnote{For example,
a value of $\BR(B\to X_s\gamma)$ somewhat lower than the SM NLO prediction
has recently stimulated a revision~\cite{KN} of the theoretical prediction,
now in perfect agreement with the experimental data.}.
At the time being, it does not look implausible to think that what the experiments see
is a SM effect, at least predominantly.


This being said, it remains a relevant question if the observed $\epsilon'$ could be
produced by supersymmetric effects
in  generic and motivated supersymmetric scenarios.
This question we intend to study in this paper.

\section{Supersymmetric flavour effects}
Supersymmetry can generate flavour and CP violation effects
orders of magnitude above the experimental bounds
(with generic soft terms) or --- at the contrary ---
get rid of any new flavour and CP-violating effect
(if soft terms are mediated by flavour blind interactions, like gauge interactions,
at not too high energies).
There are however two generic situations
where supersymmetry gives rise to new effects around the
experimental bounds:
\begin{itemize}
\item `12/3' effects: if the sparticles of the first two generations are degenerate,
while third generation sparticles have a different mass,
12/3 mixing angles comparable to the CKM ones
are not excluded and generate a variety of signals.
Appropriately weighting possible experimental
progress and theoretical evidence,
we can try to order the signals according to their interest.
Few of them appear rather promising
$$\mu\to e \gamma \hbox{ (together with other $\mu\to e$ transitions)},\qquad
d_e, \qquad d_N,\qquad
\epsilon$$
while it seems less likely (but not impossible) that effects in
$$\epsilon',\qquad
\tau\to\mu\gamma,\qquad
\Delta m_B,\qquad
b\to s\bar{s}s,\qquad
b\to s \gamma,\qquad
K\to \pi\nu\bar{\nu}$$
show up at an interesting level.

\item `1/2' effects: while generic mass matrices
of $L$eft handed and of $R$ight handed sfermions lead to unacceptably large effects
(so that we assume flavour degeneracy of the first two families of sfermions),
a generic structure of $A$ terms gives $LR$ mixing terms
$A_{ij}\lambda_{ij}$
that do not unavoidably give too large effects.
We will try to establish a hierarchy of the resulting signals
also in this case.
\end{itemize}
Both the 12/3 and the 1/2 effects are reasonably motivated by flavour physics:
for example a U(2) flavour symmetry~\cite{U2} can be invoked to
force the degeneracy among the first two generations of sfermions.
This symmetry does not restrict the third generation sfermions
nor force the universality of the $A$ terms, so that
both type of effects are present in this case.

Both type of effects may be present in unified theories
with supergravity-type mediation of soft terms.
The 12/3 effects are already there in unified theories with `minimal'
Yukawa couplings:
even starting with universal soft terms above the unification scale,
a 12/3 mass splitting between the sfermions is
generated by the unified
top Yukawa coupling~\cite{12/3}.
The 1/2 effects require non universal $A$ terms as an initial condition\footnote{Even
small Yukawa couplings generated by operators with a more complex gauge
structure than the one of the minimal Yukawa interactions (for example
by non-renormalizable operators
involving the fields that break the unified group)
are in general not enough to produce non universal $A$ terms by
RGE effects, unlike what is claimed in~\cite{1/2}.
We thank A. Pomarol and R. Rattazzi for reminding us of this point.
See~\cite{GUTdecoupling} for a discussion of decoupling effects
at the GUT scale.},
unless one complicates the structure of the theory in a non trivial way.

The 1/2 flavour violating effects arise as follows.
We assume that the Yukawa matrices for the first two generations contain small
non-diagonal elements: for example in the $d$ quark sector we parametrize them as
\begin{equation}\label{eq:d1d2}
{\lambda\over \lambda_s}= \bordermatrix{&\dq_{1L} &\dq_{2L}\cr 
\dq_{1R}&\epsilon_{dd}& \epsilon_{sd}\cr 
\dq_{2R}&\epsilon_{ds} & 1},\qquad
{A\lambda\over\lambda_s}= \bordermatrix{&\tilde{\dq}_{1L} &\tilde{\dq}_{2L}\cr 
\tilde{\dq}_{1R} &A_{11} \epsilon_{dd} & A_{21}\epsilon_{sd}\cr
\tilde{\dq}_{2R}& A_{12}\epsilon_{ds} &A_{22}},\qquad
\end{equation}
with similar expressions for leptons and u-quarks.
The Yukawa matrix is approximately diagonalized by $R_{12}(\theta)$ rotations
as $\lambda_{\rm diag}=R_{12}(\epsilon_{sd}) \lambda R_{12}^T(\epsilon_{ds})$.
In the ($d,s$) basis where $\lambda_{\rm diag}=\diag(\lambda_d,\lambda_s)=
\diag(|\epsilon_{dd}-\epsilon_{sd}\epsilon_{ds}|,1)\lambda_s$ is diagonal,
$A\lambda$ still has
non diagonal complex terms
\begin{equation}\label{eq:Alambda}
A \lambda=\lambda_s\bordermatrix{&\tilde{d}_L &\tilde{s}_L\cr
\tilde{d}_R &\pm  [A_{11}\epsilon_{dd}+(A_{22}-A_{12}-A_{21})
\epsilon_{sd}\epsilon_{ds}] &
\pm(A_{21}-A_{22}) \epsilon_{sd}\cr
\tilde{s}_R & (A_{12}-A_{22}) \epsilon_{ds} & A_{22} }
\end{equation}
where $\pm$ is the sign of $\epsilon_{dd}-\epsilon_{sd}\epsilon_{ds}$.
For example, if $\epsilon_{dd}=0$ and $|\epsilon_{sd}|=|\epsilon_{ds}|$,
the 12 entry is proportional to $\sqrt{m_d m_s}$
and $\epsilon_{sd}\approx \sqrt{m_d/m_s}$ gives the dominant contribution to the Cabibbo angle.

\medskip

It has been observed in~\cite{MM} that
CP violating 1/2 effects 
generated by non universal and complex $A$-terms can give rise
to a supersymmetric contribution to  $\epsilon'$
as large as the observed one.
One has to note, however, that if the $A$ terms were complex to start with,
one would also expect a phase in the other soft breaking parameters
(the $B$ term, or --- in the most used phase convention --- the $\mu$ term)
which would not generate $\epsilon'$.
It is well known that  electron and neutron EDMs give strong constraints
on the phases of the $\mu$ term, but do not disfavour a maximal phase
in the $A$ terms~\cite{dipoliSoft}.

\medskip

It is interesting, however, that flavour and CP violating 1/2 effects can be obtained
with real soft terms
provided that
\begin{itemize}
\item[(1)] the Yukawa couplings $\lambda_{ij}$ are complex
(as usually assumed in order to get a complex phase in the CKM matrix);
\item[(2)] the $A$ terms have the form $A_{ij}\lambda_{ij}$ with
$A_{ij}$ real and non-universal;
\item[(3)] there is no negligible entry in the 1,2 sector of the Yukawa matrix.
\end{itemize}
More precisely the condition (3) means that no element of the matrix can be set to zero
without affecting the masses in a relevant way.

In this situation, flavour and CP-violating 1/2 effects arise in the following way.
Using only redefinitions of the phases of the $d_{iL}$, $d_{iR}$
superfields it is possible to choose a basis in which only one element of the
1,2 Yukawa matrix is complex.
It is convenient to choose the only complex entry to be the 11 one
($\epsilon_{dd}$ in\eq{d1d2}).
It is now possible to obtain approximately diagonal Yukawa matrices
by performing real rotations in the $L$eft and $R$ight 12 sectors.
These rotations leave non diagonal elements in the $A\lambda$ terms if the $A$ terms
are non universal as in eq.\eq{Alambda}.
At this point we have a diagonal Yukawa matrix with a complex $\lambda_d$ element, which has
a different phase from the  11 element of the $A\lambda$ matrix, see eq.\eq{Alambda}.
With a phase redefinition of the $d_R$ field
we can reach the `usual' basis where $\lambda_d$ is real and positive.
In this `usual' basis the $dd$ and the $ds$ entries 
of the $A\lambda$ matrix are complex.
The same procedure can be repeated in the lepton sector.

\section{Typical hierarchy of the 1/2 signals}
We now want to study the expected hierarchy of the 1/2 signals: in particular
if this scenario can be motivated
as a supersymmetric origin of the observed value of $\epsilon'$.
The diagonal $A$-terms give rise, as usual, to electron and neutron EDMs.
The non-diagonal entries give rise to $\epsilon'$ (in the $d$ sector), but also to
$\mu\to e\gamma$ (in the lepton sector).

Only penguin operators are generated by 1/2 effects
(so that the supersymmetric effects in $\epsilon$, $\Delta m_B$,
$K\to \pi\nu\bar{\nu}$ are negligible).
However the hierarchy between the `penguin effects'
($\mu\to e \gamma$, $d_e$, $d_N$, $\epsilon'$, and,
if the analysis is extend to the third generation, $b\to sss$, $b\to s\gamma$, $\tau\to\mu\gamma$)
is not substantially different from the case of 12/3 effects.
The main difference is that hadronic effects are not suppressed
by possibly significant radiatively induced GIM cancellations~\cite{HFV}.

We perform our analysis in
a simple motivated case chosen in order to enhance the relative importance of $\epsilon'$
with respect to the other signals that we study.
We assume that
\begin{enumerate}
\item each Yukawa matrix has vanishing 11 entry
and comparable 12 and 21 entries;

\item the $A_{ij}$ terms are complex and non universal, while the $\mu$ term is real.

\end{enumerate}
With stronger simplifying assumptions on the $\lambda$ and $A\lambda$ matrices
we would restrict our analysis to very particular situations.
For example $\epsilon'$ vanishes for antisymmetric or symmetric matrices
with vanishing 11 element\footnote{We thank Andrea Romanino 
and Hitoshi Murayama for having pointed
out an error on this in a previous version of the paper},
while the electric dipoles vanish for hermitian matrices.
The zero in the 11 position implies 1/2 mixing angles in the $d$-quark sector equal to
$\epsilon_{ds}\approx \epsilon_{sd}\approx \sqrt{m_d/m_s}\approx 0.23$,
larger than the 1/2 mixing angles $\epsilon_{\mu e}\approx \epsilon_{e\mu}\approx 
(m_e/m_\mu)^{1/2}\approx 0.07$  between leptons.
Having made these favourable (but reasonable) choices,
we will find that a dominant effect in $\epsilon'$
is somewhat disfavoured, but nevertheless possible.

\begin{figure}[t]
\begin{center}
\begin{picture}(17.7,8)
\putps(0,0)(0,0){fEpsdN}{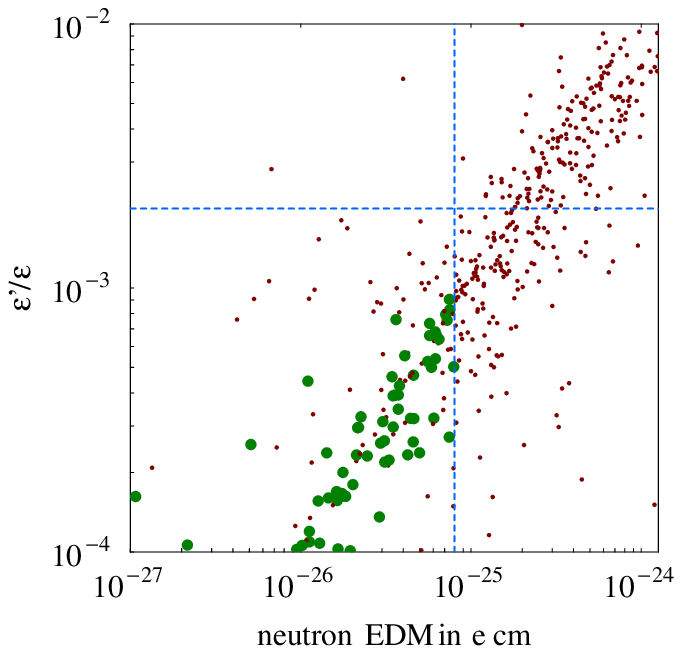}\Red
\putps(8,0)(9,0){fEpsde}{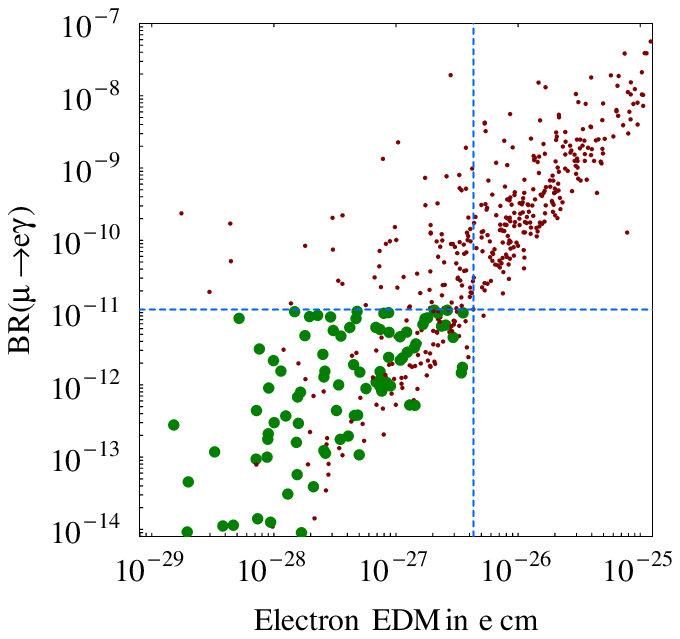}\Red\Black
\end{picture}
\caption[SP]{\em Scatter plot of
values of $(d_N,\epsilon'/\epsilon)$ (fig.\fig{scatter}a)
and of $(d_e,\BR(\mu\to e\gamma))$ (fig.\fig{scatter}b)
as generated by supersymmetry with non universal complex
$A$ terms and in the flavour model chosen in order to maximize
the relative importance of $\epsilon'$ (see text).
Sampling points excluded by a too large $d_N$, $d_e$ or $\BRmueg$
are denoted with a small dot \Red$\cdot$\Black{},
while allowed points are denoted with a bigger dot \Green$\bullet$\Black{}.
\label{fig:scatter}}
\end{center}\end{figure}

1/2 effects generate magnetic and chromo-magnetic penguins.
We give in the appendix the expressions for the relevant penguin
amplitudes and list here simple approximate expressions for the signals
\begin{eqnsystem}{sys:stime}
{\hbox{BR}(\mu\to e\gamma)\over1.1~10^{-11}} &\approx &
0.1\frac{\TeV^4}{m_{\tilde{\ell}}^4}
\frac{|A_{\mu e} \epsilon_{\mu e}|^2+|A_{e\mu} \epsilon_{e\mu}|^2}{2M_N^2   m_e/m_\mu}\\
{d_e\over 4.3~10^{-27}\ecm} &\approx& 0.1 \frac{\TeV^2}{m_{\tilde{\ell}}^2}\Im \frac{A_e^*+\mu\tan\beta}{M_N}\\
{d_N\over 0.8~10^{-25}\ecm} &\sim& 4  \frac{\TeV^2}{m_{\tilde{q}}^2}\Im \frac{A_{u,d}^*+\mu\tan\beta}{M_3}\\
\label{eq:eps'approx}
{\epsilon'/\epsilon\over 2~10^{-3}} &\sim& \frac{B}{2} \frac{\TeV^2}{m_{\tilde{q}}^2}
\Im\frac{\epsilon_{sd}^* A_{sd}^*-\epsilon_{ds}A_{ds}}{2M_3\sqrt{m_d/m_s}}
\end{eqnsystem}
In eq.\eq{eps'approx} the factor $B$,
precisely defined in the appendix in eq.\eq{matrixelement},
parametrizes the uncertainty on the relevant $K\to 2\pi$ matrix element.
It is reasonable to expect it to lie in the range $0.3\circa{<}B\circa{<}3$.
In our analysis we will employ the value $B=3$.
A similar uncertainty affects the prediction for the neutron EDM,
computed as described in the appendix.
We have defined $\lambda_s\epsilon_{ds}A_{ds}$ and
$\lambda_s\epsilon_{sd}A_{sd}$ 
as the off-diagonal components of the
$A\lambda$ matrix in\eq{Alambda},
and similarly for leptons.
The approximate expressions are not valid for $M_i\to 0$.
As discussed in the appendix, we have maybe slightly overestimated $\epsilon'$
and omitted a possibly dominant contribution to the neutron EDM
from the $s$-quark EDM~\cite{dNs}
(this contribution is not present in the case of real non-universal $A$ terms).

The approximate expressions indicate that
\begin{itemize}

\item The value of $d_N/\epsilon'$ generated by 1/2 effects
does not depend strongly
on the sparticle spectrum
and is very close to the value generated by 12/3 effects
(eq.~18 in~\cite{HFV}).
With a matrix element for $\epsilon'$ slightly larger than the one we have used, or
with a small cancellation between the various contributions to $d_N$,
the supersymmetric contribution to $\epsilon'$
can be as large as the observed value
without conflicting with the bound from neutron EDM.
Unless for some reason the imaginary part of $A\lambda_{11}$ in\eq{Alambda}
is suppressed, a supersymmetric $\epsilon'$ implies
that $d_N$ is very close to the experimental bounds
$$d_N=(-3\pm5)10^{-26}\ecm~~\cite{dNexp1990},\qquad
d_N=(1.9\pm5.4)10^{-26}\ecm~~\cite{dNexp1999}.$$

\item
The bound on $\BRmueg$~\cite{MEGA} requires sleptons heavier than $\sim1\TeV$
(maybe helped by a light neutralino and/or by a small leptonic $A$ term).

\item The bound from the electron EDM is not very significant.
We remember that we have taken $\mu$ real,
while the $A$ terms have large complex phases.
A complex $\mu$ term would have been a more efficient source of
EDMs (because a large $\mu\tan\beta$ is suggested in minimal models by the following considerations:
a too low $\tan\beta\circa{<}2$ tends to give a too light higgs,
and a large $\mu$-term can compensate the large RGE gluino correction to EWSB)
without enhancing $\epsilon'$.
\end{itemize}
These results are confirmed by a numerical analysis, as shown in fig.\fig{scatter}.
To study the correlation between hadronic and leptonic observables,
we assume universal sfermion masses and universal gaugino masses at the unification scale.
We do not require universal $A$ terms at the GUT scale, but
we assume that the $A_{ij}$ terms that multiply the d Yukawa couplings
are unified with the ones that multiply the leptonic Yukawa couplings.
We scan these parameters in the following way:
$$m_0,|\mu_0|,M_5,|B_0|,|A_{22}|,|A_{12}|,|A_{21}|=(0\div 1)m_{\rm SUSY}.$$ 
where $m_{\rm SUSY}$ is a dimensionful parameter,
and $(0\div 1)$ means a random number between 0 and 1 (different for each parameter).
The phases of the $A$ parameters are extracted as $\exp [2\pi i (0\div 1)]$,
while the $\mu$ term is taken real (positive or negative).
The value of $m_{\rm SUSY}$ is determined by imposing that the minimum of the
MSSM higgs potential be at its physical value.
This scanning procedure~\cite{naturalness} generates
a sample of spectra with density inversely proportional to the `fine-tuning'
(i.e.\ configurations that require unlikely cancellations are more rare).

We restrict our analysis to the few sampling spectra
that still satisfy the LEP and Tevatron bounds on sparticle masses,
that can be roughly summarized as
$$m_\chi>91\GeV, \qquad
m_h\circa{>}85\GeV,\quad
m_{\tilde{\ell}}>80\GeV\qquad
M_3,m_{\tilde{q}}\circa{>}(180\div 260)\GeV.$$
Consequently in most of the cases the lightest chargino has mass close
to $100\GeV$ (the distribution of sparticle masses
is plotted in fig.~4 of~\cite{naturalness}).
This procedure automatically takes into account the possibility
of cancellations between different contributions
to the observables we are studying.

Fig.s\fig{scatter} clearly show that a significant contribution
to $\epsilon'$, although possible, is not likely {\em if we maintain
the described correlation between quarks and leptons.}
This, however, may not be necessarily the case.

\medskip

As said, we have restricted our analysis to a simple 
structure of Yukawa matrix.
To see what would happen in a different case, it is sufficient to rescale
our final results by appropriate factors.
For example it is possible to increase the value of $\epsilon'/d_N$ without generating
a too large contribution to the Cabibbo angle by assuming higher values of
$\epsilon_{ds}$.
For example $\epsilon_{ds}\sim 1$ can be motivated by flavour symmetries.
In unified models, however, this also implies $\epsilon_{\mu e}\sim 1$,
enhancing $\BRmueg$ faster than $\epsilon'$,
see eq.s\sys{stime}
(and, depending on the value of neutrino masses, giving too large
$\nu_e\to \nu_\mu$ oscillations). 
This alternative choice seems thus
more problematic than the one we have assumed.

\smallskip

The pattern of effects generated by complex Yukawa matrices and real non universal $A$ terms
is qualitatively similar (although there are more unknown parameters
that cannot be fixed in a simple motivated way),
so that we do not show the corresponding plot.

\medskip

If this set of effects were observed,
it will be interesting to know if they are due to 1/2 effects or to 12/3 effects
(or to something else).
Along the lines of~\cite{de/dN},
the value $d_e/d_N$ could give an indirect discrimination of the source.
If sparticles will be produced at accelerators, one can check if there is the 12/3
non-degeneracy among sfermions (for example between $\tilde{e}_R$ and $\tilde{\tau}_R$) characteristic of 12/3 effects.
It might also be possible to observe flavour violation in sfermion interactions:
the amplitudes for the 1/2 effects that we have studied
will be suppressed by factors like $m_\mu/m_{\tilde{\mu}}$,
while amplitudes for 12/3 effects would only be suppressed by not very small CKM-like mixings.

\section{Conclusions}
Present theoretical uncertainties in the determination
of the $\epsilon'$ parameter do not allow a firm conclusion
on the plausibility of explaining the measured value in the SM.
In turn, the interpretation of the experimental result as a (main) supersymmetric
effect appears highly speculative,
at the time being.

It is nevertheless true that a plausible mechanism exists~\cite{MM}
able to generate an effect in $\epsilon'$.
A way to confirm such an effect would be to discover a neutron
electric dipole just below the present bounds.

Even with favourable structures of the Yukawa couplings, a simple minded correlation
between quarks and leptons makes the present limit on $\mu\to e\gamma$ highly constraining
if not excluding at all the possibility of a large supersymmetric contribution to $\epsilon'$
in motivated models.
This is even more true for the 12/3 effect than for the 1/2 effect,
because of GIM-like cancellations distinguishing quarks and leptons in the former case.
The correlation between quarks and leptons can, however, be less straightforward or even
not exist at all.
In a truly unified field theory we believe that $\mu\to e\gamma$ is anyhow the
most sensitive probe of supersymmetric flavour effects.
As such, the interpretation of $\epsilon'$ as a (mainly) supersymmetric effect
is a further motivation to look for flavour violations in the leptonic sector.

\appendix
\setcounter{equation}{0}
\renewcommand{\theequation}{\thesection.\arabic{equation}}

\section{Loop effects}

1/2 effects generate magnetic and chromo-magnetic penguins.
If we parametrize the coefficients of these operators as
$$A(f_L\to f'_R\gamma)
\bigg[\frac{1}{2}(\bar{f}_L\gamma_{\mu\nu}F_{\mu\nu} f'_R) + {\rm h.c.}\bigg]+
A(f_L\to f'_R G)
\bigg[\frac{1}{2}(\bar{f}_L\gamma_{\mu\nu}T^a G^a_{\mu\nu} f'_R) + {\rm h.c.}\bigg]+
(L\leftrightarrow R)
$$
the values of the penguin amplitudes $A$ are
\begin{eqnarray*}
A(d\to d\gamma)&=&-m_d 2 c_3 q_d\frac{e\alpha_3}{4\pi }
\Im \frac{A_d^*+\mu \tan\beta}{M_3^3}P'_{BI}(r_{d3})\\
A(d\to dG)&=&m_d\frac{g_3\alpha_3}{4\pi M_3^2}
\Im \frac{A_d^*+\mu \tan\beta}{M_3^3}[-(2c_3-c_8)P'_{BI}(r_{d3})-c_8 P'_{FI}(r_{d3})]\\
A(s_L\to d_R G)&=&c_8 m_s\epsilon_{sd}^* \frac{g_3\alpha_3}{4\pi}
 \frac{A_{sd}^*}{M_3^3}[ P'_{FI}(r_{d3})-P'_{BI}(r_{d3})/9]\\
A(\mu_L\to e_R \gamma) &=&-q_e m_\mu \epsilon_{\mu e}^*
\frac{e\alpha_Y}{4\pi} A_{\mu e}^*\sum_{n=1}^4 \frac{H_{n\tilde{B}}(H_{n\tilde{B}}+\cot \theta_{\rm W} H_{n\tilde{W}_3})}{M_{N_n}^3}P'_{BI}(r_{eN_n})
\end{eqnarray*}
with analogous expressions for the $u$ dipole and for
$s_R\to d_L$ and $\mu_R\to e_L$ transitions. 
The flavour diagonal amplitudes ($f=f'$) coincide with the EDMs:
$A(f\to f\gamma)=d_f$ and
$A(f\to f\gamma)=d_f^{\rm QCD}$.
Here $c_3=4/3$ and $c_8=3$ are QCD factors
and $r_{fi}\equiv m_{\tilde{f}}^2/M_i^2$ is a ratio
between sfermion and gaugino/higgsino squared masses.
The $P$ functions are defined e.g. in eq.~(B.6) in ~\cite{b->sss}.
If sfermions of opposite chirality are degenerate,
$P'(r)$ is the derivative of $P(r)$.
In general $P'(r_{fi})$ stands for $[P(r_{f_Li})-P(r_{f_Ri})]/(r_{f_Li}-r_{f_Ri})$.
The neutralino mixing matrix $H$ diagonalizes the neutralino mass matrix $M_N$ as $H M_N H^T$.

The QCD renormalization corrections to the electric and chromo-electric dipoles
have been computed integrating the RGE equations~\cite{dqRGE}\footnote{We have recomputed
the $\gamma$ coefficient for $d_q$, since its sign differs from the one in various recent phenomenological analyses
but it agrees
with the original computation~\cite{dqRGE}.}
$$\frac{d}{d\ln\mu}\pmatrix{d_q\cr d_q^{\rm QCD}}=\frac{\alpha_3}{4\pi} \gamma^T\cdot\pmatrix{d_q\cr d_q^{\rm QCD}},
\qquad\gamma= \pmatrix{8/3 &0\cr 32 eq_q/3g_3 & (-29+2n_f)/3}$$
from $\mu\sim M_Z$ down to a QCD scale at which $\alpha_3(\mu)\sim 4\pi/6$.
The factor $n_f$ is the number of active quark flavours and $q=\{u,d\}$.
The neutron EDM $d_N$ has been computed in terms of the $u$ and $d$ quark EDMs renormalized at the QCD scale
using the naive quark model approximation and
taking into account chromo magnetic dipoles using the naive dimensional analysis~\cite{dq->dN}:
$$d_N\approx \frac{1}{3}(4d_d-d_u)+\frac{e}{4\pi}(4 d_d^{\rm QCD}-d_u^{\rm QCD}).$$
We are omitting a possibly dominant contribution from the $s$-quark EDM~\cite{dNs}
(not present with real non-universal $A$ terms).

The magnetic penguin operators that contribute to $\epsilon'$ renormalize in the same way as
the dipole operators.
It is however convenient to include a $g_3$ in the operator (rather than in its coefficient), so that
the anomalous dimension  of the chromo magnetic dipole that gives rise to $\epsilon'$ becomes $\gamma=+4/3$.
Its matrix element has been
taken from~\cite{EpsPMatrixElement}
\begin{equation}
\label{eq:matrixelement}
\bAk{\pi\pi~I=0}{g_3~\bar{s}\gamma_{\mu\nu}T^a G^a_{\mu\nu} \frac{1\pm\gamma_5}{2} d}{K_0}\approx
\pm\sqrt{\frac{3}{2}}\frac{11}{4}\frac{F_K^2}{F_\pi^3}\frac{m_K^2m_\pi^2}{m_s}B,\qquad
\frac{\epsilon'}{\epsilon}=\frac{\omega\bAk{\pi\pi~I=0}{{\cal H}_{\rm eff}}{K_0}}{\sqrt{2}\epsilon~ \Re A_0}
\end{equation}
where, with our normalization of the meson states,
$F_\pi=131\MeV$, $F_K=160\MeV$, $\Re A_0=3.3~10^{-7}\GeV$ and $\omega=1/22$.
The $B$ factor, computed at the leading order in the chiral quark model, is $B\approx 0.3$~\cite{EpsPMatrixElement}.
However $B$ could be larger since the next orders are not suppressed by the $m_\pi^2$ factor.
In our numerical analysis we have employed the value $B=3$.

The $\mu\to e \gamma$ branching ratio is given by
$$\BRmueg=(m_\mu^3/16\pi)
(|A(\mu_L\to e_R\gamma)|^2+|A(\mu_R\to e_L\gamma)|^2)/\Gamma(\mu).$$

\end{document}